\documentclass[12pt,a4paper,fleqn]{article}
\usepackage[textheight=23cm,textwidth=16cm]{geometry}
\usepackage{amsmath}
\usepackage{graphicx}
\usepackage{cite}
\usepackage[american]{babel}

\usepackage{here}
\begin{document}

\begin{center}

{\LARGE\bf
 New Benchmark Set of Transition-Metal Coordination Reactions for the Assessment of Density Functionals
}

\vspace{2cm}

{\large
Thomas Weymuth$^{\rm a}$,
Erik P.~A.~Couzijn$^{\rm b}$,
Peter Chen$^{\rm b, }$\footnote{E-Mail: chen@org.chem.ethz.ch}, \\
and Markus Reiher$^{\rm a, }$\footnote{E-Mail: markus.reiher@phys.chem.ethz.ch}
}\\[4ex]

$^a$ ETH Zurich, Laboratorium f\"ur Physikalische Chemie, \\
Vladimir-Prelog-Weg 2, 8093 Zurich, Switzerland

$^b$ ETH Zurich, Laboratorium f\"ur Organische Chemie, \\
Vladimir-Prelog-Weg 2, 8093 Zurich, Switzerland

\vspace{7cm}

\vfil

\end{center}

\begin{tabbing}
Date:   \quad \= March 19, 2014 \\
Status:       \> submitted to \textit{J.~Chem.~Theory Comput.} \\
\end{tabbing}

\newpage

\begin{abstract}

We present the WCCR10 data set of ten ligand dissociation energies of large cationic transition metal complexes for the assessment of 
approximate exchange--correlation functionals. We analyze nine popular functionals, namely BP86, BP86-D3, B3LYP, B3LYP-D3, 
B97-D-D2, PBE, TPSS, PBE0, and TPSSh by mutual comparison and by comparison to experimental gas-phase data measured with well-known 
precision. The comparison of all calculated data reveals a large, system-dependent scattering of results with nonnegligible 
consequences for computational chemistry studies on transition metal compounds. Considering further the comparison with experiment, the 
non-empirical functionals PBE and TPSS turn out to be among the best functionals for our reference data set. The deviation can 
be lowered further by including Hartree--Fock exchange. Accordingly, PBE0 and TPSSh are the two most accurate functionals 
for our test set, but also these functionals exhibit deviations from experiment by up to 50\,kJ\,mol$^{-1}$ for individual reactions. 
As an important result we found no functional to be reliable for all reactions. Furthermore, for some of the ligand dissociation 
energies studied in this work, dispersion corrections yield results which increase the deviation from experiment.
This deviation increases further if structure optimization including dispersion corrections is performed. 
Finally, we compare our results to other benchmark studies and highlight that the performance assessed for 
different density functionals depends significantly on the reference molecule set chosen.

\end{abstract}

\newpage

\section{Introduction}

Density functional theory (DFT) is an important tool in computational quantum chemical research\cite{koch01,dyks05,rev_mod_phys_1999_71_1253}. 
This is due to the fact that DFT offers a good compromise between accuracy and computing time, which allows to investigate 
large molecules with hundreds of atoms. The essential ingredient of DFT, the exact exchange--correlation functional, is not 
known. During the last decades, efforts were made to develop better and better approximations to the exact exchange--correlation 
functional (by increasingly advanced analytic ans\"atze and improved reference data sets for their parametrization)\cite{aip_conference_proceedings_2001_577_1}. 
As a result, a plethora of (approximate) density functionals is now at our disposal. Unfortunately, there is no truly rigorous 
way to systematically develop such approximations, and their reliability is assessed in statistical analyses by comparison 
to reference data sets. As a consequence, a given density functional might provide reasonable or even excellent results in 
some application, but completely fail in another one.

Comparing the values of one or several chemical properties calculated with different density functionals with accurate reference 
data, obtained either experimentally or by quantum chemical methods of controllable accuracy (e.g., coupled cluster\cite{helg00}), 
provides a means to measure the quality of these density functionals, at least with respect to the properties studied. Pople 
and coworkers were among the first to realize the importance of benchmark studies, which they applied to assess the quality 
of the so-called G1 and G2 composite approaches\cite{j_chem_phys_1989_90_5622,j_chem_phys_1991_94_7221}. In 1981, they 
proposed a comparatively small database containing the atomization energies of 31 small molecules such as LiH, CH$_4$, 
CO$_2$, and H$_2$NNH$_2$\cite{j_chem_phys_1989_90_5622}. This database has been continuously extended; in 1997, G2/97 was 
presented which comprised already 148 molecules\cite{j_chem_phys_1991_94_7221,j_chem_phys_1997_106_1063}, followed by 
the G3/99 test set with 222 molecules in 1999\cite{j_chem_phys_2000_112_7374}. In 2005, the latest revision of this 
Pople benchmarking database was presented\,---\,G3/05 includes a total of 454 reference data points\cite{j_chem_phys_2005_123_124107}. 

More recently, Truhlar and coworkers assembled a large collection of databases including atomization energies, ionization 
potentials, electron and proton affinities, barrier heights, anharmonic vibrational zero point energies, saddle point 
geometries, and a broad range of binding energies (covering systems with hydrogen bonds, charge-transfer complexes, dipole 
interactions, and weakly interacting systems)\cite{j_phys_chem_a_2003_107_1384,j_phys_chem_a_2003_107_3898,j_phys_chem_a_2003_107_8996,j_phys_chem_a_2004_108_2715,j_phys_chem_a_2005_109_2012,j_chem_theory_comput_2005_1_415,j_phys_chem_a_2005_109_4388,j_phys_chem_a_2005_109_11127,theor_chem_accounts_2008_120_215,j_phys_chem_c_2008_112_6860,j_chem_theory_comput_2009_5_324}. 
This broad range of different properties shall assure a rather complete picture of the performance of a given density 
functional. In a similar spirit, Grimme and Korth proposed ``mindless'' benchmarking relying on a large library of 
automatically generated ``artificial'' molecules, which should remove any chemical biases\cite{j_chem_theory_comput_2009_5_993}. 
Goerigk and Grimme incorporated this test set into the GMTKN24 database covering main group thermochemistry, kinetics, and 
noncovalent interactions\cite{j_chem_theory_comput_2010_6_107}, which was recently extended to form the so-called GMTKN30 
database\cite{j_chem_theory_comput_2011_7_291}.

A drawback of all of these databases is the fact that the coverage of large transition metal complexes is rather poor. This 
deficiency can be significant for practical purposes as coordination chemistry in the condensed phase usually involves 
complex, spatially extended ligand environments. By contrast, a compact small diatomic like Cr$_2$ with unsaturated valencies 
can be a true challenge for electronic structure methods, but it is no prototypical system for coordination chemistry.

While the Grimme databases are specifically set up for main group chemistry and do therefore not cover any transition metals, 
the Truhlar databases include some small transition metal compounds like Ag$_2$, CuOH$_2^+$, Fe(CO)$_5$, ferrocene, and bis(benzene)chromium. Jiang 
{\it et al.}~developed a database with the heats of formation of 225 molecules containing 3d-transition metals\cite{j_phys_chem_a_2012_116_870}. 
However, most of these molecules are also rather small. In addition, many of the experimental reference data have a rather 
large uncertainty, which might result in wrong conclusions as to the performance of a given density functional. Very recently, 
Zhang {\it et al.}~therefore selected the 70 molecules which are estimated to have experimental uncertainties in their 
enthalpies of formation of at most 2\,kcal\,mol$^{-1}$\cite{j_chem_theory_comput_2013_doi} and tested no less than 
42 density functionals on that database. However, this reduced dataset again contains only very small compounds; furthermore, 
only a minority of them contains organic ligands. Hughes and Friesner assembled a rather large database of spin-splittings of 
57 octahedral first-row transition metal complexes\cite{j_chem_theory_comput_2011_7_19}. Furche and Perdew also presented a 
database of 3d transition metals, including reaction energies. Again, also this database includes only very small compounds 
(and reactions such as the dissociation of the vanadium dimer)\cite{j_chem_phys_2006_124_044103}. By contrast, {\it large} 
organometallic compounds are key players in important fields such as homogeneous catalysis\cite{duca12}. It is often not 
possible to draw conclusions from quantum chemical calculations on small complexes for larger compounds. 
Furthermore, dispersion effects have a much larger influence in bigger complexes. Therefore, a database of reliable values 
for important chemical properties like ligand dissociation energies for a variety of large transition metal complexes is 
highly desirable. 

For such large complexes, no truly reliable {\it ab initio} computational data are currently available. Therefore, for
a functional investigation of large complexes that shall reach beyond a mutual comparison of popular functionals, we have to resort to experimental 
data. Ideally, these data should not only be accurate but also obtained in the gas phase, such that strong (environmental) 
intermolecular interactions like solvent effects can be excluded. Some of us accurately measured ligand dissociation energies of 
a broad range of large transition metal complexes in the gas phase by means of tandem mass spectrometry. The experimental 
ligand dissociation energies were obtained by injecting the intact complex into a mass spectrometer and letting it subsequently 
collide with an inert noble gas such as argon or xenon. This collision led to dissociation of a ligand, and the dissociation 
cross sections were measured as a function of the nominal collisional energy. The actual value of the ligand dissociation energy 
was then determined by Monte Carlo simulations based on a rather involved model of the entire dissociation process involving 
the ligand dissociation energy as a fit parameter\cite{j_phys_chem_a_2007_111_7006}. This procedure assures a sufficient and determinable 
accuracy of the ligand dissociation energies. Here, we present a database of ten selected ligand dissociation energies from Refs.~\cite{organometallics_2005_24_1907,inorg_chem_2007_46_11366,j_am_chem_soc_2008_130_4808,j_am_chem_soc_2011_133_8914,chemphyschem_2010_11_1002,j_am_chem_soc_2010_132_13789,chem_eur_j_2010_16_5408,kobylianski_preparation} and investigate the performance of nine widely used density functionals.

This work is organized as follows: In section \ref{sec:compmet}, we introduce our new database of coordination energies (which we will abbreviate 
in the following as WCCR10), the density functionals investigated, and the computational benchmarking procedure. The results 
are discussed in section \ref{sec:results} before providing a conclusion together with an outlook in section \ref{sec:conclusion}.

\section{Computational Methodology}
\label{sec:compmet}

\subsection{WCCR10 Ligand Dissociation Energy Database of Large Transition Metal Complexes}

\begin{figure}[H]
 \begin{center}
   \includegraphics[scale=0.5]{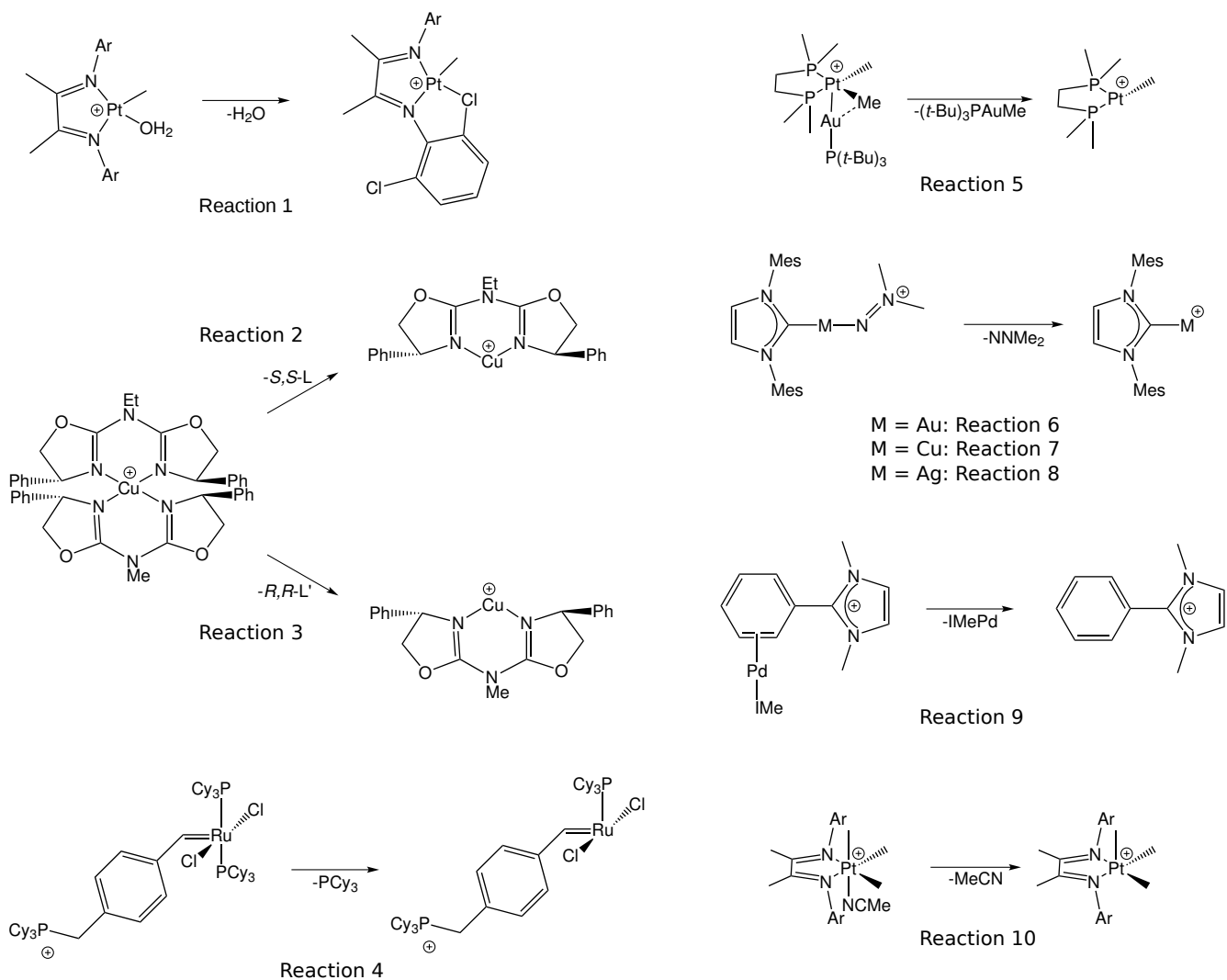}
 \end{center}
 \caption{\label{fig:reactions}\small Overview of the ten reactions of the WCCR10 database. The aromatic substituent abbreviated 
as ``Ar'' is 2,6-C$_6$H$_3$Cl$_2$.}
\end{figure}

Our database consists of the ten dissociation reactions depicted in Fig.~\ref{fig:reactions}. These reactions were selected out 
of all such reactions that have been published so far with the experimental methodology mentioned in the introduction\cite{organometallics_2005_24_1907,inorg_chem_2007_46_11366,j_am_chem_soc_2008_130_4808,j_am_chem_soc_2011_133_8914,chemphyschem_2010_11_1002,j_am_chem_soc_2010_132_13789,chem_eur_j_2010_16_5408}. Specific care was taken to achieve a 
good balance of different transition metals and different ligand types. Hence, duplication of similar ligand dissociation reactions 
has been avoided. All molecules in this database feature a closed-shell electronic structure. All transition metal complexes 
are charged, which is an implication of the fact that the ligand dissociation energies have been determined by mass spectrometry.

Table \ref{tab:reference} lists the experimental ligand dissociation energies for the ten reactions depicted in Fig.~\ref{fig:reactions}. 
The error bounds were estimated in the original literature from approximately 10 to 20 regressions with the L-CID program\cite{j_phys_chem_a_2007_111_7006} 
for three independent data sets. Furthermore, a lab-frame energy uncertainty of 0.15\,eV was assumed\cite{organometallics_2005_24_1907,inorg_chem_2007_46_11366,j_am_chem_soc_2008_130_4808,j_am_chem_soc_2011_133_8914,chemphyschem_2010_11_1002,j_am_chem_soc_2010_132_13789,chem_eur_j_2010_16_5408}. These experimental ligand dissociation energies are based on reaction rates derived within approximate 
Rice--Ramsperger--Kassel--Marcus (RRKM) theory and can therefore be considered to be free energies at 0 Kelvin. Therefore, 
these energies have to be compared to computed total electronic energies which are corrected for their zero-point vibrational 
energies.

Reactions 4 and 5 feature a twofold degeneracy of the reaction paths, i.e., there are two equivalent pathways to product formation, involving either of the two phosphine ligands for reaction 4 and either of the two platinum-bound methyl groups for reaction 5. However, the reported experimental reaction energies\cite{j_am_chem_soc_2008_130_4808,j_am_chem_soc_2011_133_8914} were determined with a version of L-CID\cite{j_phys_chem_a_2007_111_7006} that did not take into account this degeneracy in the microcanonical RRKM treatment of the rates. As a result, the reaction rates as a function of collision energy have been overestimated by a factor of two, which slightly affects both the kinetic shift treatment and the fitted reaction energies and indirectly, the other fitting parameters. This typically leads to a minor underestimation of the reaction energies by $<$\,3\,kJ\,mol$^{-1}$ (this corresponds well to the 2\,kJ\,mol$^{-1}$ difference obtained from the Eyring equation, which applies to canonical ensemble rate theory). Whereas this error is within the experimental uncertainty of the reaction energies, we decided for the WCCR10 set to refit the original experimental data sets with a version of L-CID that does take into account the twofold reaction path degeneracy\cite{chem_eur_j_2010_16_5408,kobylianski_preparation}. The new data obtained for reactions 4 and 5 are given in Table \ref{tab:reference} (see the Supporting Information for further details).

\begin{table}[H]
 \renewcommand{\baselinestretch}{1.0}
 \renewcommand{\arraystretch}{1.0}
 \caption{\label{tab:reference}\small Experimental reference values for the ten ligand dissociation energies in the WCCR10 database.}
 \begin{center}
  \begin{tabular}{l r r} \hline \hline
reaction & E$_{\rm ref}$ / (kJ\,mol$^{-1}$) & Ref. \\
\hline
1        & 120.2\,$\pm$\,2.7$^a$          & \cite{organometallics_2005_24_1907} \\
2        & 204.4\,$\pm$\,5.0              & \cite{inorg_chem_2007_46_11366} \\
3        & 207.3\,$\pm$\,2.9              & \cite{inorg_chem_2007_46_11366} \\
4        & 136.7\,$\pm$\,3.8              & this work, \cite{j_am_chem_soc_2008_130_4808} \\
5        & 192.5\,$\pm$\,5.0              & this work, \cite{j_am_chem_soc_2011_133_8914} \\
6       & 211.9\,$\pm$\,6.7              & \cite{chemphyschem_2010_11_1002,j_am_chem_soc_2010_132_13789} \\
7       & 218.2\,$\pm$\,5.4              & \cite{chemphyschem_2010_11_1002,j_am_chem_soc_2010_132_13789} \\
8       & 194.4\,$\pm$\,8.4              & \cite{chemphyschem_2010_11_1002,j_am_chem_soc_2010_132_13789} \\
9        & 162.7\,$\pm$\,4.2              & \cite{chem_eur_j_2010_16_5408} \\
10        & 102.6\,$\pm$\,2.5              & \cite{kobylianski_preparation} \\
\hline
\hline
  \end{tabular}
   \renewcommand{\baselinestretch}{1.0}
   \renewcommand{\arraystretch}{1.0}
 \end{center}
$^a$ Obtained by taking the arithmetic mean of two independent values which are obtained with different collisional gases. 
With argon, a value of 121.6\,$\pm$\,3.8\,kJ\,mol$^{-1}$ is found, while with xenon, the ligand dissociation energy is found 
to be 118.7\,$\pm$\,3.8\,kJ\,mol$^{-1}$. Moreover, these two values were obtained with the program {\sc Crunch}\cite{j_chem_phys_1997_106_4499}. 
In this case, the error bounds are estimated from the reproducibility across data sets and fitting assumptions (also here, 
a lab-frame energy uncertainty of 0.15\,eV was assumed).
\end{table}

\subsection{Computational Details}
\label{sec:compdet}

In this study, seven different density functionals are investigated, namely BP86\cite{bp86a,bp86b}, B3LYP (including 20\% of 
Hartree--Fock exchange)\cite{phys_rev_b_1988_37_785,j_phys_chem_1994_98_11623}, B97-D-D2\cite{j_comput_chem_2006_27_1787}, 
TPSS\cite{phys_rev_lett_2003_91_146401}, TPSSh (10\% Hartree--Fock exchange)\cite{j_chem_phys_2003_119_12129,j_chem_phys_2004_121_11507}, PBE\cite{phys_rev_lett_1996_77_3865,phys_rev_lett_1997_78_1396}, and PBE0 (25\% Hartree--Fock exchange)\cite{j_chem_phys_1999_110_5029}. 
Note that we adopt the special designation ``B97-D-D2'' for the functional which is usually just referred to as ``B97-D'' 
in order to make clear that this functional is a reparametrized version of the original ``B97'' functional, and that Grimme's 
second generation dispersion correction (``D2'')\cite{j_comput_chem_2006_27_1787} should be used in conjunction with this 
functional. Furthermore, for BP86 and B3LYP we investigate Grimme's third generation dispersion correction\cite{j_chem_phys_2010_132_154104}, 
denoted as BP86-D3 and B3LYP-D3, respectively. This collection of nine functionals includes generalized gradient approximation 
(GGA, BP86, PBE), meta-GGA (TPSS), and hybrid functionals (B3LYP, TPSSh, PBE0).

For the benchmark study, the program package {\sc Turbomole} version 6.3.1 in its shared-memory parallelized implementation 
was employed\cite{turbomole}. We made an attempt to fully optimize all structures in $C_1$ symmetry (i.e., without any 
structural constraints) with each of the above-mentioned density functionals. Hence, the ligand dissociation energies were in 
general {\it not} obtained by simple single-point calculations (apart from a few exceptions\,---\,to be discussed below). 
This methodology allows us to eliminate any errors stemming from an inconsistent description of the molecular structure.

In order to minimize the basis set superposition error, Ahlrichs' very large quadruple-$\zeta$ basis with two sets of 
polarization functions (def2-QZVPP) was employed at all atoms\cite{phys_chem_chem_phys_2005_7_3297} (note that for transition 
metal complexes, the basis set size may have a much larger effect on the accuracy than previously assumed\cite{j_comp_chem_2013_34_870}). 
For all non-hybrid functionals, advantage was taken of the resolution-of-the-identity (RI) approximation with the corresponding 
auxiliary basis sets\cite{aux-def2-universal}. Scalar relativistic effects were taken into account for all second- and third-row 
transition metal elements by means of Stuttgart effective core potentials\cite{theor_chim_acta_1990_77_123} as implemented 
in {\sc Turbomole}. In all structure optimizations, the maximum norm of the Cartesian geometry gradient was converged to 
10$^{-4}$\,hartree/bohr. If not stated otherwise, all total electronic energies were converged to within 10\textsuperscript{$-$7}\,hartree. 
For all other program settings, the default values were found to be suitable for our purpose.

The reaction energies have to be corrected for the zero-point vibrational energies (ZPE). These have been computed with BP86 
within the harmonic approximation. For all other density functionals, the ZPE resulting from BP86 was taken since our tests 
showed (see also below) that the different density functionals yield very similar ZPEs. For all molecules except for the 
modified Grubbs catalyst of reaction 4 (see Fig.~\ref{fig:reactions}), the vibrational analysis was carried out with the 
methodology just described, but using the fine grid ``m4'' and electronic energies converged to 10\textsuperscript{$-$8}\,hartree. 
The total number of basis functions for the modified Grubbs catalyst (having 174 atoms), however, was too large for  
{\sc Turbomole's aoforce} module. Therefore, in this case the vibrational analysis was carried out with {\sc Snf} 5.0.0 as 
provided with {\sc MoViPac} 1.0.0\cite{j_comput_chem_2012_33_2186}. This program implements a semi-numerical differentiation 
scheme which allows to construct the Hessian matrix as numerical first derivatives of analytical geometry gradients of the 
total electronic energy\cite{snf}. The geometry gradients have been computed with the serial version of {\sc Turbomole} 6.3.1. 
For the numerical differentiation, a 3-point central difference formula with a step size of 0.01\,bohr was utilized.

The entire benchmark is computationally very expensive, as very accurate calculations with tight self-consistent field 
convergence criteria and huge basis sets on large molecules had to be conducted. A typical single-point computation required 
a total computing time of about 100 days (featuring only a small number of about 20 SCF iterations), corresponding to a wall 
time of a few days when employing a parallel version. Consequently, the full structure optimizations required several months 
to complete. Unfortunately, it turned out that for some combinations of reactions and density functionals (namely, reactions 
2, 3, and 4 for B3LYP, B3LYP-D3, TPSSh, and PBE0, and reaction 10 for B3LYP-D3) the structure optimizations were not feasible 
for some of the molecules involved. However, we found for all non-dispersion-corrected functionals that the individual 
structures agree very well with each other (see section \ref{sec:structures}). Therefore, for the above-mentioned reactions and functionals 
we decided to include data obtained by single-point calculations (taking the necessary precautions in our analysis). 
The structures resulting from optimizations with dispersion-corrected functionals, however, are rather different from each 
other. Therefore, we cannot present any data for reactions 2, 3, 4, and 10 obtained with the B3LYP-D3 
functional. However, these restrictions do not compromise the complete and coherent picture that emerges for the functionals 
investigated.

\section{Results}
\label{sec:results}

\subsection{Structures}
\label{sec:structures}

\begin{figure}[H]
 \begin{center}
   \includegraphics[scale=0.1]{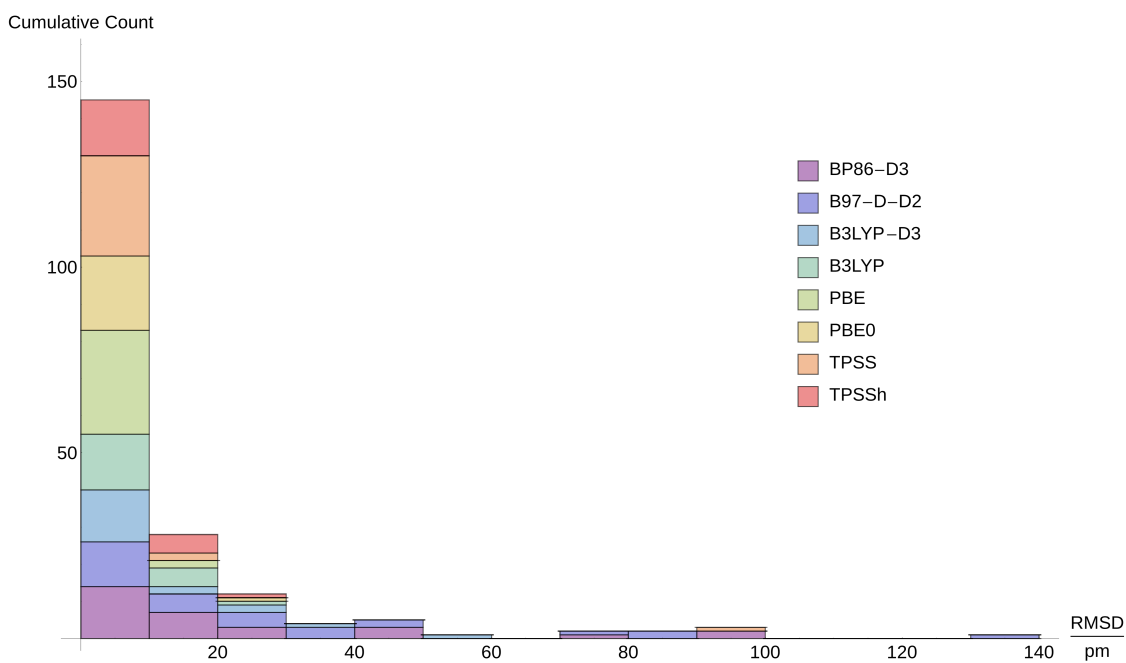}
 \end{center}
 \caption{\label{fig:rmsd}\small Distribution of RMSD values for all optimized structures, taking the structures resulting 
from BP86 as reference and using a bin size of 10\,pm. Note that we plot a cumulative count on the abscissa (i.e., the {\it total} 
number of molecules having a RMSD between 0 and 10\,pm is 145).}
\end{figure}

The first crucial step is to analyze the effect that the individual functionals have on the molecular structures optimized 
(Cartesian coordinates of all optimized structures can be found in the supporting information). To this end, we compare the 
structures resulting from different density functionals to each other in terms of the root mean square deviation (RMSD) 
between the atomic positions of a given structure and the corresponding BP86-optimized structure (the BP86 functional is known 
to yield reasonable structures\cite{koch01}). The resulting RMSD values are shown in Fig.~\ref{fig:rmsd}. We see that about 
91\% of all structures have an RMSD below 30\,pm. Visual inspection of these structures shows that they are almost identical 
to the BP86 reference. Especially the PBE optimized structures agree almost perfectly with the corresponding BP86 structures 
(we note that the exchange part of PBE is similar to B88, while the correlation part of PBE differs from P86\cite{bp86a,bp86b,phys_rev_lett_1996_77_3865}). 
This similarity of structures suggests that a full structure optimization might not be necessary for each individual density 
functional. Instead, single-point calculations may be sufficient. To investigate this aspect further, we calculated the ligand 
dissociation energy of reaction 9 for selected density functionals not only from a full structure optimization, but also by means 
of single points on BP86 structures. These data are given in Table \ref{tab:approximations} (last two columns). We see that 
the individual coordination energies agree very well with each other, with deviations hardly exceeding 0.5\,kJ\,mol$^{-1}$. 
This finding allows us to present ligand dissociation energies also for the few cases where a full structure optimization was not 
possible (cf., section \ref{sec:compdet}).

\begin{figure}[H]
 \begin{center}
   \includegraphics[scale=0.5]{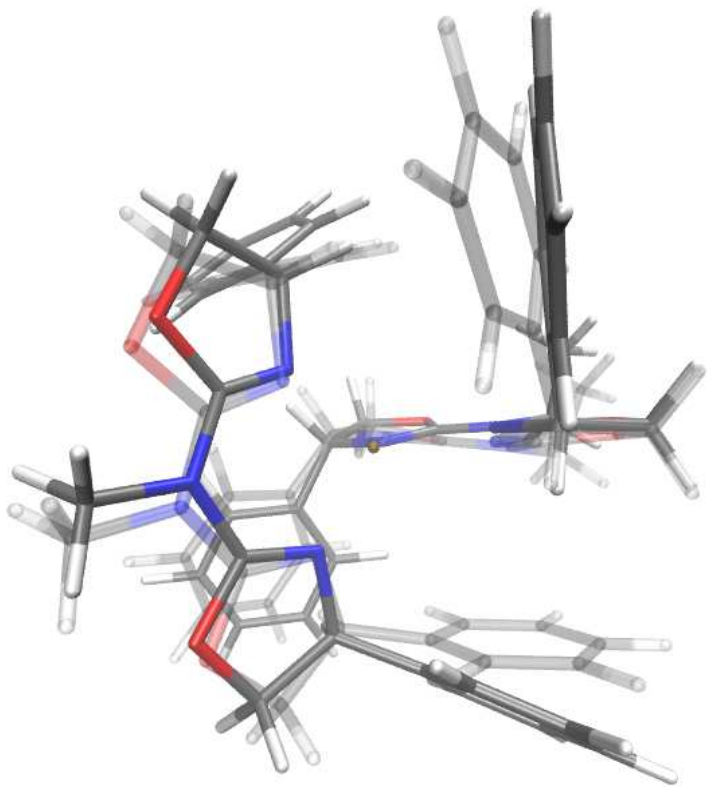}
 \end{center}
 \caption{\label{fig:dispersion}\small Overlay of the structures of the reactant complex of reactions 2 and 3 resulting from optimization 
with BP86 (solid structure) and BP86-D3 (translucent structure). It is clearly visible that the phenyl substituents align in 
an almost coplanar fashion with the five-membered rings when dispersion corrections are considered.}
\end{figure}

However, there are also a few structures with large RMSD values (i.e., larger than 30\,pm); interestingly, all except one are 
obtained with a dispersion-corrected functional. Typically, the structures with the largest 
RMSD values are the transition metal complexes of reactions 2 and 3 (see Fig.~\ref{fig:reactions}). When analyzing the 
structures of these complexes we found that the overall molecule ``contracts'' when dispersion corrections are taken into 
account (as one would expect), and the phenyl substituents adopt a stacked, almost coplanar, conformation with respect 
to the five-membered rings of these complexes (see Fig.~\ref{fig:dispersion}). Thus, large intramolecular dispersion effects 
are responsible for the large differences observed here. For the reactant complex of reaction 4, we also found rather large
structural changes upon taking dispersion interactions into account. These changes are discussed in the Supporting Information.

\begin{figure}[H]
 \begin{center}
   \includegraphics[scale=0.1]{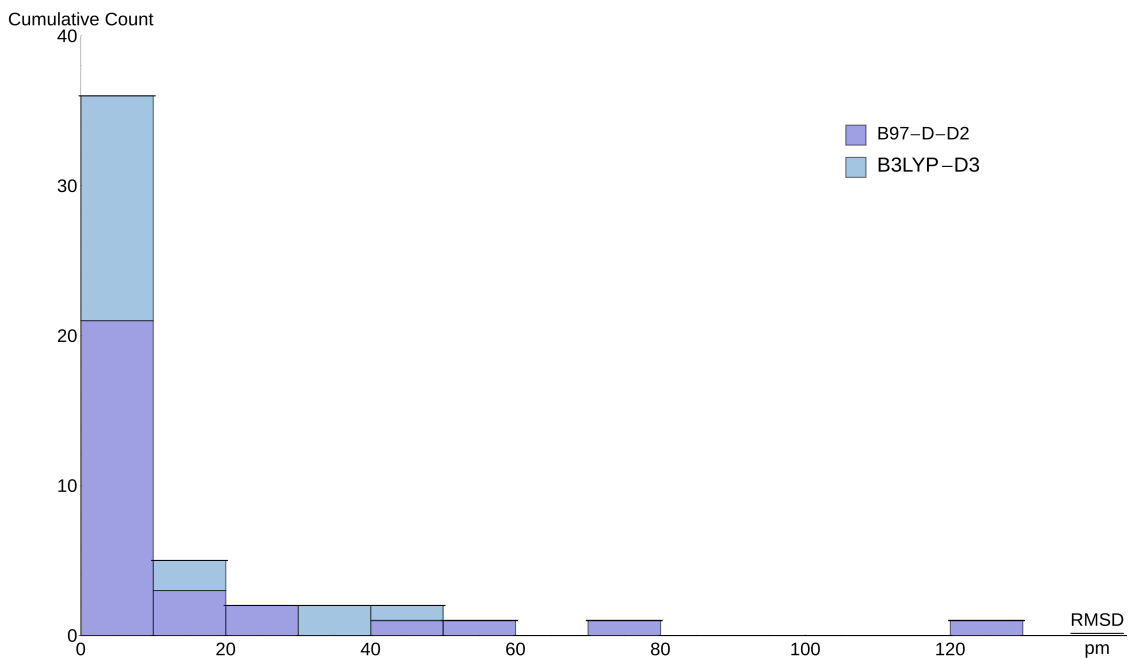}
 \end{center}
 \caption{\label{fig:rmsd-dispersion}\small Distribution of RMSD values for the structures optimized with the dispersion-corrected 
functionals B3LYP-D3 and B97-D-D2, taking the structures resulting from BP86-D3 as reference and using a bin size of 10\,pm.}
\end{figure}

It is not surprising that the structures resulting from dispersion-corrected functionals are rather different from their 
BP86 reference, since the latter does not take any long-range dispersion effects into account. However, we found that the 
structures optimized with B3LYP-D3 and B97-D-D2 deviate also rather strongly from the ones obtained with BP86-D3 (see Fig.~\ref{fig:rmsd-dispersion}). 
Therefore, in the case of dispersion-corrected density functionals, a full optimization is necessary in order to avoid any 
artificial errors stemming from an unoptimized and thus for a given functional not well-defined structure. For example, the 
ligand dissociation energy of reaction 3 is calculated to be 297\,kJ\,mol$^{-1}$ when the structures are optimized with BP86-D3. 
When only single points on the BP86 structures are considered instead, the reaction energy is found to be 256\,kJ\,mol$^{-1}$, 
i.e., there is a significant difference of 41\,kJ\,mol$^{-1}$. At this point, we should emphasize that it is not possible 
to determine how reliable either structure is as there are no experimental data available with which one may compare. However, the fact that 
the structures obtained with different dispersion-corrected functionals differ significantly from each other demands caution 
when applying dispersion corrections in structure optimizations with certain density functionals. We should stress already at 
this point that we do not intend to favor one type of dispersion correction over another, but aim at a detailed survey
of some of the popular and thus highly used present-day functionals.

Many of the complexes studied in this work feature moieties (often the methyl substituents) that have very low rotational 
barriers (for example, the rotation around the P--Au bond of reactant complex 5 has a barrier of roughly 3.4\,kJ\,mol$^{-1}$). The 
optimization algorithm may not necessarily find the exact minimum with respect to these rotations, even though tight convergence 
criteria were employed. However, since these rotational barriers are so low, this should have a 
negligible effect on the overall ligand dissociation energies. Furthermore, in most cases these ligands with low rotational barriers are present 
both in the reactant as well as the product complexes, such that one can expect a cancellation of errors to occur when 
calculating the reaction energies.

\subsection{Zero-Point Vibrational Energies}

For a sensible comparison with the experimental reference energies of the WCCR10 reference set, we have to correct the total 
electronic energies by their respective zero-point energies (ZPE), which requires a full vibrational analysis. However, for 
the large transition metal complexes investigated in this work, this is hardly feasible for every density functional. Therefore, 
we decided to calculate the ZPE only with BP86, and to subsequently use this ZPE for all other density functionals as well. 
The unscaled harmonic frequencies obtained with BP86 are known to be in good agreement with experimental fundamental frequencies, 
which must be due to a fortunate cancellation of errors\cite{inorg_chem_2002_41_6928,j_chem_phys_2003_118_7215}. Still, we 
tested the validity of this approximation at the example of the ligand dissociation energy of reaction 9 
(cf., Fig.~\ref{fig:reactions}). These data are given in Table \ref{tab:approximations}.

\begin{table}[H]
 \renewcommand{\baselinestretch}{1.0}
 \renewcommand{\arraystretch}{1.0}
 \caption{\label{tab:approximations}\small Zero-point vibrational energies (ZPE) and ligand dissociation energies ($\Delta$E$_i$) 
of reaction 9 for selected density functionals. $\Delta$E$_1$ is the ligand dissociation energy calculated from a full structure 
optimization to a local minimum on the respective potential energy surface, followed by vibrational analysis to compute the 
ZPE. $\Delta$E$_2$ represents the ligand dissociation energy calculated from a structure optimization after which, however, no 
vibrational analysis has been carried out\,---\,instead, the ZPE is always calculated with BP86. $\Delta$E$_3$ finally is 
the ligand dissociation energy calculated from single points, using both the structures and the ZPE from BP86. All values are 
given in kJ\,mol$^{-1}$.}
 \begin{center}
  \begin{tabular}{l r r r r} \hline \hline
functional & ZPE & $\Delta$E$_1$ & $\Delta$E$_2$ & $\Delta$E$_3$ \\
\hline
BP86      & 0.68 & 137.16        & 137.16        & 137.16         \\
BP86-D3   & 0.66 & 163.70        & 164.89        & 164.35         \\
PBE       & 0.82 & 146.09        & 146.29        & 146.91         \\
TPSS      & 0.48 & 143.39        & 143.07        & 143.37         \\
B3LYP     & 1.46 & 107.93        & 109.23        & 109.37         \\
\hline
\hline
  \end{tabular}
   \renewcommand{\baselinestretch}{1.0}
   \renewcommand{\arraystretch}{1.0}
 \end{center}
\end{table}

For this selected reaction, we calculated the ligand dissociation energies from a structure optimization followed by a full 
vibrational analysis. These data are denoted as $\Delta$E$_1$ in Table~\ref{tab:approximations}. This computational reference 
has to be compared with $\Delta$E$_2$, which represents the ligand dissociation energies obtained from a structure optimization, 
but taking the ZPE from the BP86 calculation. 

We learn from Table \ref{tab:approximations} that the ZPE corrections vary between 0.5\,kJ\,mol$^{-1}$ in the case of 
TPSS and 1.5\,kJ\,mol$^{-1}$ for B3LYP. Clearly, these variations are tiny compared to those found in the corresponding 
ligand dissociation energies, which vary over a range of almost 60\,kJ\,mol$^{-1}$. In fact, we find that the values of $\Delta$E$_2$ 
agree well with the corresponding values of $\Delta$E$_1$. Therefore, we conclude that the $\Delta$E$_2$ methodology adopted 
for our benchmark is sufficiently accurate.

\subsection{Ligand Dissociation Energies}

\begin{figure}[H]
 \begin{center}
   \includegraphics[scale=0.1]{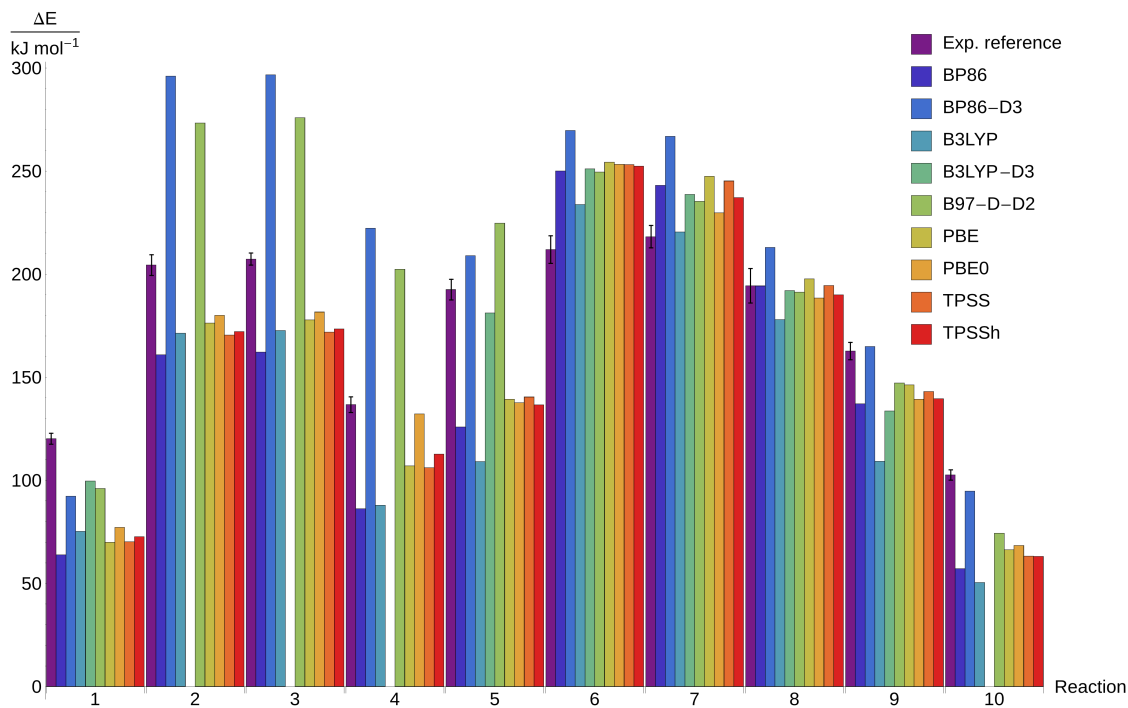}
 \end{center}
 \caption{\label{fig:energies}\small Experimental and calculated ligand dissociation energies of the ten benchmark reactions studied. 
For the B3LYP-D3 functional, not all data are available (see also section \ref{sec:compdet}). For the experimental data, the 
black bars denote the error bounds.}
\end{figure}

We are now in a position to analyze the ligand dissociation energies of the individual reactions as calculated with the nine 
different density functionals. These energies are plotted together with their experimental reference values in Fig.~\ref{fig:energies}. 
While for some reactions all density functionals yield very similar energies (e.g., reaction 8), there is an enormous spread 
for other reactions, most notably for reactions 2 and 3. The energies calculated with the dispersion-corrected functionals 
BP86-D3 and B97-D-D2 (recall that no data are available for B3LYP-D3) deviate strongly not only from the values of all other 
density functionals, but also from the experimental reference (the same is true for reaction 4 but not for all other reactions). 
Obviously, in these cases, the dispersion corrections do not lead to an improved performance, but, on the contrary, make the 
result worse compared to the performance of the non-dispersion-corrected reference BP86. In some cases, the ligand dissociation energy is overestimated 
by more than 90\,kJ\,mol$^{-1}$ (in case of reaction 2, the BP86 energy is 161\,kJ\,mol$^{-1}$, while the corresponding 
BP86-D3 value is 296\,kJ\,mol$^{-1}$; the experimental reference is 204.4\,kJ\,mol$^{-1}$). We therefore conclude that 
dispersion corrections in the form proposed 
by Grimme do not necessarily lead to an improved performance for coordination energies. Very recently, Kobylianskii {\it et al.}~reported similar findings for the gas-phase Co--C bond energies of adenosylcobinamide 
and methylcobinamide\cite{j_am_chem_soc_2013_135_13648}. We should mention here that in a previous study, Jacobsen and Cavallo 
found that the inclusion of dispersion interactions led to worse phosphane substitution energies for a range of transition metal 
complexes\cite{chemphyschem_2012_13_562}. In that study, this effect was, however, attributed to an incomplete treatment of 
solvent effects\cite{chemphyschem_2012_13_562}, which was also confirmed by Grimme\cite{chemphyschem_2012_13_1407,chemphyschem_2012_13_1405}. 
Since we compare to gas-phase experimental data, solvation effects are not present and thus cannot be a reason for the larger 
deviation of dispersion-corrected density functionals from experiment when compared to their generic parent functionals (compare Ref.\ \cite{schm13}). We 
will investigate this issue in some more detail in section \ref{sec:dispersion} below.

\begin{figure}[H]
 \begin{center}
   \includegraphics[scale=0.1]{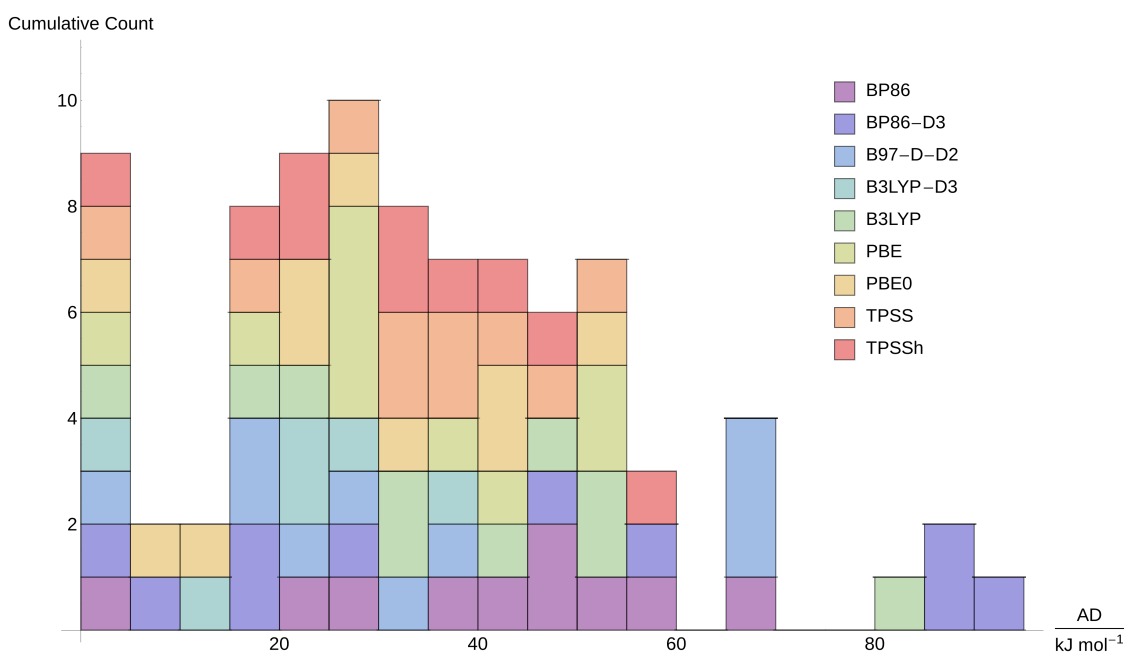}
 \end{center}
 \caption{\label{fig:errors}\small Distribution of the absolute deviations (AD) of the ligand dissociation energies as calculated 
with the individual density functionals from their experimental reference values.}
\end{figure}

Most of the (dispersion-interaction-free) calculated 
ligand dissociation energies deviate between 15 and 55\,kJ\,mol$^{-1}$ from their experimental reference. However, in some cases 
an almost perfect agreement is found (for example, reaction 8 calculated with BP86 and reaction 7 calculated with B3LYP). 
The distribution of the absolute deviations from the experimental reference values is shown in Fig.~\ref{fig:errors}. 
%We can hardly establish any trends. 
Based on these results no functional may be clearly preferred over the others.

\begin{table}[H]
 \renewcommand{\baselinestretch}{1.0}
 \renewcommand{\arraystretch}{1.0}
 \caption{\label{tab:errors}\small Mean absolute deviations (MAD) and largest absolute deviations (LAD) of the ligand 
dissociation energies as calculated with the individual functionals (first two columns), as well as MAD values for selected functionals 
obtained in three previously published studies. All values are given in kJ\,mol$^{-1}$.}
 \begin{center}
  \begin{tabular}{l r r r r r} \hline \hline
functional & MAD            & LAD         & MAD  & MAD  & MAD  \\
          & (this work)    & (this work)       & (study 1)\cite{j_chem_theory_comput_2009_5_324} & (study 2)\cite{j_chem_theory_comput_2013_doi} & (study 3)\cite{j_chem_phys_2006_124_044103} \\
\hline
BP86      & 39.58           & 66.57            & n/a           & n/a           & 43.05         \\
BP86-D3   & 44.60           & 91.62           & n/a           & n/a           & n/a           \\
B3LYP     & 39.08           & 83.42            & 46.15         & 20.90         & 50.16         \\
B3LYP-D3  & 20.48$^a$       & 39.15$^a$        & n/a           & n/a           & n/a           \\
PBE       & 31.83           & 53.14            & 16.89         & 25.50         & 45.14         \\
B97-D-D2  & 36.09           & 68.95            & n/a           & n/a           & n/a            \\
TPSS      & 32.93           & 52.06            & 17.97         & n/a           & 42.64          \\
TPSSh     & 31.99           & 55.93            & 15.97         & 17.56         & 40.55          \\
PBE0      & 26.88           & 54.81            & 12.58         & 19.23         & n/a            \\
\hline
\hline
  \end{tabular}
   \renewcommand{\baselinestretch}{1.0}
   \renewcommand{\arraystretch}{1.0}
 \end{center}
$^a$ Note that for B3LYP-D3, no data are available for reactions 2, 3, 4, and 9. For reactions 2, 3, and 4, the other dispersion-corrected 
density functionals (BP86-D3 and B97-D-D2) have particularly large errors. We might therefore expect that the values presented in this 
Table would be significantly worse for B3LYP-D3 if these data could be taken into account.
\end{table}

\begin{figure}[H]
 \begin{center}
   \includegraphics[scale=0.11]{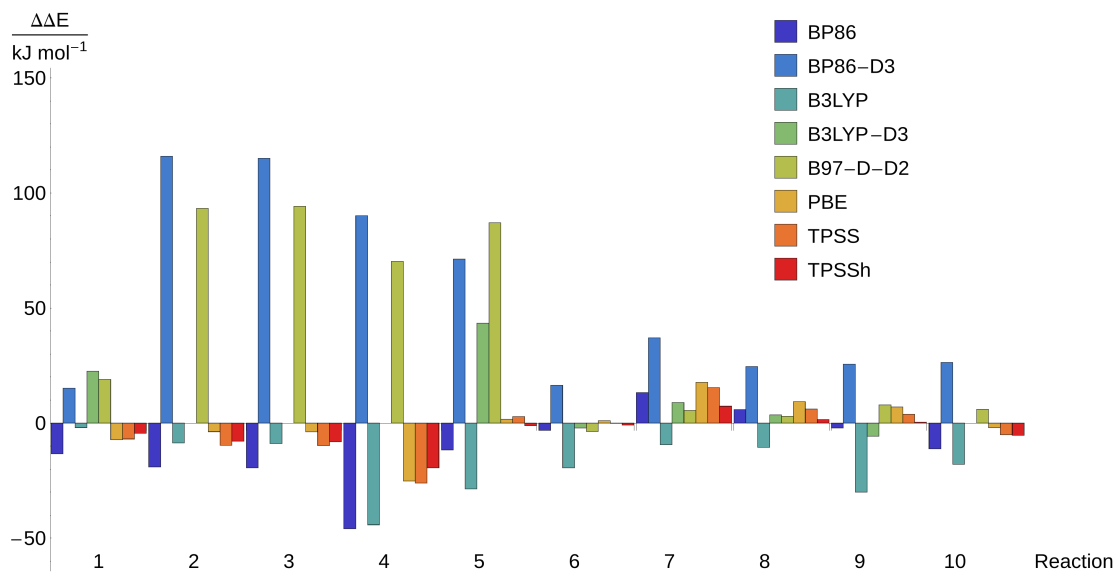}
 \end{center}
 \caption{\label{fig:energies2}\small Deviations of dissociation energies calculated with different functionals from those calculated with PBE0, which has been
chosen as a reference.}
\end{figure}

By investigating the mean absolute deviations (MADs) of the individual functionals (see Table \ref{tab:errors}), we see that 
most functionals have MADs of about 30\,kJ\,mol$^{-1}$. The MAD of the dispersion-corrected functionals BP86-D3 and 
B97-D-D2 is clearly larger, which is due to the above-mentioned significant errors in the case of reactions 2, 3 and 4. We note again 
that there is no data available for these reactions for B3LYP-D3, and we may expect that the MAD of the results obtained with this functional would otherwise be significantly larger.
The largest absolute deviation (LAD) from the experimental values is about 50\,kJ\,mol$^{-1}$ for most functionals. For BP86 the largest deviation is 67\,kJ\,mol$^{-1}$, 
and for B3LYP it is even as large as 83\,kJ\,mol$^{-1}$. Excluding the dispersion-corrected functionals, B3LYP performs worst 
considering both the MAD as well as the largest absolute deviation from experiment. BP86 has a comparable MAD, but its LAD is 
significantly better than in the case of B3LYP. A considerable improvement can be obtained by using the 
non-empirical PBE functional. With 32\,kJ\,mol$^{-1}$ its mean absolute deviation is about 8\,kJ\,mol$^{-1}$ lower than that 
of BP86. The LAD is improved by roughly 13\,kJ\,mol$^{-1}$, being about 53\,kJ\,mol$^{-1}$. Moving from this GGA functional to 
the non-empirical meta-GGA functional TPSS, no further improvement is obtained; both the MAD as well as the LAD are almost identical 
to those of PBE. In the case of these two functionals, admixture of Hartree--Fock exchange can further decrease the mean 
absolute deviations. In particular for PBE0, the MAD decreases to 27\,kJ\,mol$^{-1}$, which is the lowest value of all density 
functionals investigated in this work. However, adding Hartree--Fock exchange does not generally lead to a reliable description 
of the ligand dissociation energies, as exemplified by the B3LYP results. Note that in an earlier benchmark study by Fuche and Perdew\cite{j_chem_phys_2006_124_044103}, 
the TPSS and TPSSh functionals were found to be best functionals for dipole moments, reaction energies, and harmonic frequencies 
of 3d transition metal complexes. Still, the best-performing functionals in our study have errors of up to about 50\,kJ\,mol$^{-1}$ 
for certain reactions and thus no functional performs well for our complete WCCR10 set.
Even in a 'theory-only' comparison the scatter of the calculated results is significant: Fig.~\ref{fig:energies2} 
provides a comparison of deviations in dissociation energy (with the PBE0 functional chosen as a reference).

\subsection{A Closer Inspection of Dispersion Corrections}
\label{sec:dispersion}

Considering again Fig.~\ref{fig:energies}, we note that the energies of reactions 2, 3, and 4 calculated with dispersion-corrected 
functionals deviate from both the experimental values and the values calculated with all other density functionals.
It is therefore appropriate to investigate this aspect in more detail.

Grimme's dispersion corrections were parametrized for 
pairs of atoms in specific valence states within a reference molecule. As the dispersion coefficients are calibrated against 
accurate results on neutral systems, this might lead to too large dispersion coefficients for cationic systems. Since there are 
no anionic counterions present in our study (the dispersion coefficients of which might then be estimated to be too small), there 
cannot be any cancellation of errors as discussed in Ref.~\cite{j_chem_phys_2010_132_154104}.

One might argue that another possible cause for the deviation of dispersion-corrected energies observed here could be rooted in the large size of 
the complexes studied (note that the dispersion-corrected functionals perform particularly badly for reactions 2, 3, and 4, which 
involve the largest compounds of the WCCR10 test set). It has been found by Tkatchenko and von Lilienfeld that for large, bulky 
molecules (and even more so for condensed-phase systems), three-body interactions must not be neglected\cite{phys_rev_b_2008_78_045116,j_chem_phys_2010_132_234109}. 
This has also been confirmed by Risthaus and Grimme\cite{j_chem_theory_comput_2013_9_1580}. While it is in principle possible 
to account for these three-body dispersion effects by means of an Axilrod--Teller--Muto term, this correction is not used by 
default in Grimme's set of dispersion corrections\cite{j_chem_theory_comput_2013_9_1580}. In order to assess their magnitude, 
we calculated the three-body correction term for the structures obtained with the standard D3 dispersion 
correction. The resulting values for this three-body correction are very similar for the functionals BP86-D3, B3LYP-D3 and 
B97-D-D2, so that we may focus on the results obtained with BP86-D3. We find that the three-body 
interaction leads in all cases to a repulsive term, thus slightly destabilizing the molecule. This 
destabilization is more pronounced the larger the molecules are. For example, in the case of the modified Grubbs catalyst 
(174 atoms; reaction 4), the three-body term amounts to a contribution of +14.0\,kJ\,mol$^{-1}$, while it is +8.9\,kJ\,mol$^{-1}$ for the 
azabox copper(I) complex of reaction 3 (90 atoms, cf., Fig.~\ref{fig:dispersion}), and only +1.6\,kJ\,mol$^{-1}$ for the aquo platinum(II) complex of reaction 
1 (42 atoms). Including the three-body correction thus leads to slightly smaller ligand dissociation energies. In the case of reaction 
3, the ligand dissociation energy is 7.5\,kJ\,mol$^{-1}$ smaller, while the ligand dissociation energy 
of reaction 4 is lowered by 5.3\,kJ\,mol$^{-1}$. Clearly, these corrections do not significantly improve the agreement with the experimental 
values and can therefore not explain the discrepancies.

However, ligand dissociation energies could be sensitive to the spatial cut-off function, 
since the reactions studied involve the formation of a {\it short} metal--ligand bond (dispersion is difficult to model at 
short distances\cite{dalton_trans_2011_40_11176}). In order to investigate this issue, we calculated the ligand dissociation energy 
of reaction 3 with BP86-D3 in combination with the Becke--Johnson (BJ) damping function (the standard damping function of 
the D3 correction is often called zero-damping, as the dispersion interaction between two atoms approaches zero if the internuclear 
distance approaches zero). With BP86-D3-BJ the reaction energy is calculated to be 298\,kJ\,mol$^{-1}$, 
while it is found to be 297\,kJ\,mol$^{-1}$ with BP86-D3. Therefore, the damping function has only a minor effect on the overall 
ligand dissociation energy of reaction 3. It is thus not likely that it can account for the discrepancies observed in 
this work. This result is in line with the work by Grimme and coworkers who found that the damping function in general has a 
negligible effect on the results (such as ligand dissociation energies)\cite{j_comput_chem_2011_32_1456}.

In order to better understand the role of dispersion interactions on the molecular structures and ligand dissociation energies,
we calculated the reaction energy of reaction 8 also with BP86-D3 and B3LYP-D3 single-point calculations employing the BP86 and
B3LYP structures, respectively. With BP86, the reaction energy is found to be 194\,kJ\,mol$^{-1}$. When calculating this energy
with BP86-D3 at the BP86 molecular structures, a value of 208\,kJ\,mol$^{-1}$ is found, i.e., the ligand dissociation
is more endothermic by 14\,kJ\,mol$^{-1}$. When carrying out also the structure optimizations with BP86-D3, the ligand dissociation
energy is even more endothermic, namely 213\,kJ\,mol$^{-1}$. We conclude from this that the structural changes due to dispersion
interactions amount to a net effect of 5\,kJ\,mol$^{-1}$ on the reaction energy. Regarding B3LYP and B3LYP-D3, a similar picture emerges. B3LYP
yields a reaction energy of 178\,kJ\,mol$^{-1}$ for reaction 8. It is changed to 184\,kJ\,mol$^{-1}$ when dispersion interactions
are considered by means of single-point calculations on the B3LYP structures. If these interactions are also taken into account 
during the structure optimizations, the reaction energy is 192\,kJ\,mol$^{-1}$.

We also analyzed the explicit values for the C$_6$ and C$_8$ dispersion coefficients for selected reactions. For reactions 2, 3, and 4, in which the 
dispersion corrections lead to particularly large discrepancies, data is available only for BP86-D3 and B97-D-D2. The dispersion
coefficients of the D2 and D3 methods are rather different from each other, which is due to the fact that the D2 correction
includes only the (scaled) C$_6$ terms, whereas the D3 method includes (unscaled) C$_6$ and C$_8$ terms. By contrast,
considering the dispersion coefficients invoked for the calculation of the reactant complex of reaction 8 very similar values are applied
in the BP86-D3 and B3LYP-D3 calculations. Still, the many small
deviations between the dispersion coefficients for BP86-D3 and B3LYP-D3 add up to a difference of 35\,kJ\,mol$^{-1}$ in the dispersion 
energies for this compound (the total dispersion energy being
$-$215\,kJ\,mol$^{-1}$ in the case of BP86-D3 and $-$180\,kJ\,mol$^{-1}$ for B3LYP-D3). The reaction energies of reaction 8
(213\,kJ\,mol$^{-1}$ for BP86-D3 and 192\,kJ\,mol$^{-1}$ for B3LYP-D3) agree well with the experimental reference of
194\,kJ\,mol$^{-1}$\,---\,in fact, with B3LYP-D3 the reference value is almost perfectly reproduced. However, it is very intriguing
that the structures deviate more from each other (the BP86-D3 and B3LYP-D3 structures have a RMSD of 50\,pm), than those 
of the corresponding parent functionals BP86 and B3LYP (the RMSD in this case being 17\,pm).

\subsection{Comparison with Other Benchmark Studies}

It is instructive to compare our results presented above with the findings obtained in previous studies. From the plethora of 
benchmark studies published so far we select here a few which are comparable to our work in the sense that they also focus 
on transition metal compounds. The data obtained in these studies is included in Table \ref{tab:errors}. In the first study 
(which we dub ``study 1'' here), Zhao and Truhlar\cite{j_chem_theory_comput_2009_5_324} studied a model system for the Grubbs 
II catalyst. They compared the energies of important stationary points in the catalytic cycle of the metathesis reaction obtained 
with different density functionals to accurate reference energies obtained with a CCSD(T)-based composite approach. As the system 
studied in this work is very similar to reaction 4 of our work, one would expect a similar performance for the different density 
functionals. In fact, we observe from the data in Table \ref{tab:errors} the same qualitative ordering, namely, PBE0 performs 
best, while PBE, TPSS, and TPSSh are only slightly worse. B3LYP, however, is significantly worse compared to the other functionals. 
Nevertheless, we also observe some interesting differences compared to our study. B3LYP performs 10\,kJ\,mol$^{-1}$ worse in 
the work of Zhao and Truhlar, while the MADs of other functionals are approximately 10\,kJ\,mol$^{-1}$ better than in our 
benchmark.  This is in agreement with our study as we find a deviation between the experimental ligand dissociation energy of reaction 4 
and the one calculated with B3LYP of 48.7\,kJ\,mol$^{-1}$, while PBE0 reproduces this energy rather accurately, having an error 
of only 4.5\,kJ\,mol$^{-1}$ (i.e., this error is even smaller than the one found in study 1). However, the functionals TPSS and 
PBE have an absolute error of roughly 30\,kJ\,mol$^{-1}$ for reaction 4, and are thus found to be significantly worse in our 
study than in study 1.

In a second, very recent study (hereafter called ``study 2''), Zhang {\it el al.}~investigated 70 small 3d transition metal compounds 
and compared DFT results to accurate experimental data\cite{j_chem_theory_comput_2013_doi}. These compounds are different compared to 
those in our WCCR10 set. Nevertheless, the results of this study are in rather good agreement with our results, which might be an 
indication of a certain transferability of both results. The overall errors of PBE0 and PBE are approximately 7\,kJ\,mol$^{-1}$ lower compared to our findings, and 
TPSSh performs even roughly 14\,kJ\,mol$^{-1}$ better in study 2. However, it might come as a surprise that B3LYP has a MAD of only 
21\,kJ\,mol$^{-1}$ in study 2, which is in sharp contrast to our findings.

Finally, Furche and Perdew\cite{j_chem_phys_2006_124_044103} also assessed the performance of different density functionals on a 
data set of small 3d transition metal compounds. Here, we focus on their results from the analysis of 18 different reaction energies 
(such as the dissociation of V$_2$, CoH, and CrF), for which experimental data are also available. We denote this third study as 
``study 3'' in Table \ref{tab:errors}. Overall, Furche and Perdew found larger errors for the individual functionals than in our 
study, which might be due to the unsaturated valencies in the small molecules leading to peculiar electron-correlation effects. The 
qualitative ordering is, however, roughly comparable to that found in our study.

In summary, we have seen that the actual numerical values for the absolute error of a given density functional can vary significantly, 
depending on the data set employed for the tests. The qualitative ordering of different density functionals appears to be more robust 
in this respect, although also here, different test sets can lead to different conclusions. Therefore,  
when considering the performance of a given density functional one has to pay special attention to how this performance was assessed. While very large, diverse 
databases might give a good indication of the average performance of a density functional, they might under- or overestimate the 
errors of that density functional in a specific application. Likewise, a very specialized database covering only a certain class of 
compounds can hardly be considered representative for a broad area of chemistry.

\section{Conclusions}
\label{sec:conclusion}

We have presented the database WCCR10, a set of ten selected ligand dissociation energies for large transition metal complexes 
for which accurate, consistently measured experimental gas-phase data are available in the literature. We then investigated the 
performance of nine different density functionals in a mutual comparison and with respect to the experimental ligand dissociation energies. The density functionals selected 
encompass a broad range of different functional classes, namely GGA, meta-GGA, and hybrid functionals; in addition, for some 
functionals also Grimme's third generation dispersion correction was employed. The calculations were set up such that any result
observed has to be attributed to the density functional and not to other factors such as an incomplete 
basis set or numerical artifacts due to loose thresholds.

In our benchmark study, we found for some reactions that Grimme's dispersion correction yields reaction energies which overshoot
those from experiment and from results obtained without dispersion corrections. Furthermore, we 
have identified cases in which different dispersion-corrected functionals can lead to significantly different structures, whereas 
the structures resulting from the uncorrected parent functionals were in much better mutual agreement. This may not be taken as 
an argument against such dispersion corrections as the parent functional apparently plays a crucial role. Unfortunately, a definitive assessment 
of the quality of the structures is not possible since there are no reliable reference data to compare with, but our study 
indicates that care should be taken when applying dispersion corrections in structure optimizations.

Among all functionals not corrected for dispersion, B3LYP performs worst with a mean absolute deviation of about 39\,kJ\,mol$^{-1}$, 
while BP86 is a little bit better, at least in terms of the largest absolute deviation. The non-empirical density functionals PBE and TPSS have a clearly smaller error both in terms 
of mean absolute deviation and largest absolute deviation. This result is particularly appealing from a theoretical point of view, 
since these density functionals are constructed by choosing a mathematical expression that satisfies constraints which would also 
be fulfilled by the true exchange--correlation functional, instead of by simply fitting it to a reference set of chemical data. Adding a 
portion of Hartree--Fock exchange to these functionals improves their performance further. We have compared our findings to 
selected benchmark studies from the literature and found that the actual numerical values for the errors of the individual density 
functionals depend heavily on the test set chosen. The relative performance of different density functionals is\,---\,at least 
from a qualitative point of view\,---\,somewhat more robust with respect to the choice of the test set, but also here one finds 
astonishing differences. Therefore, care must be taken when utilizing such benchmark results to assess the performance of density 
functionals.

Even the best functional identified for the WCCR10 set, PBE0, has a rather large mean absolute deviation. On average, the ligand 
dissociation energies deviate by more than 20\,kJ\,mol$^{-1}$, while the maximum deviation is even as large as 55\,kJ\,mol$^{-1}$. 
As these values lie well within the range of chemically relevant energies (which are on the order of a few kJ\,mol$^{-1}$ to a 
few hundred kJ\,mol$^{-1}$), the accuracy of contemporary density functionals often cannot be expected to be sufficient for a 
given application. It is therefore of utmost importance to further analyze the density functionals available today in order to 
find possible cures to their weaknesses. Relevant for the development and benchmarking of new or improved functionals is the study of composite 
approaches\cite{theor_chem_acc_2012_131_1079}\,---\,as exemplified by the so-called correlation-consistent composite approach (ccCA method) 
of Wilson and coworkers\cite{j_phys_chem_a_2007_111_11269,j_phys_chem_a_2012_116_870,j_chem_theory_comput_2012_8_4102}, the 
Gaussian\,4 method of Curtiss and coworkers\cite{j_phys_chem_a_2009_113_5170}, the method of Peterson and coworkers\cite{j_phys_chem_a_2009_113_7861},
the benchmarking efforts by Martin and co-workers \cite{kesh14},
and a very recent approach developed by Bross {\it et al.}\cite{j_chem_phys_2013_139_094302}\,---\,or a reinvestigation of ancient
parts of present-day functionals by, for instance, explicit reconstruction from accurate (spin) densities\cite{j_chem_phys_2011_135_244102,j_chem_phys_2013_138_044111}.

\section*{Acknowledgments}

This work has been supported by the Swiss National Science Foundation SNF (projects 200020\_144458 and 200020\_137505).

\section*{Supporting Information}

Cartesian coordinates of all optimized structures investigated in this work, discussion of structural changes due to dispersion 
corrections of the reactant complex of reaction 4, discussion on the mechanism of reactions 2 and 3,  and details of the redetermination of the energies 
of reactions 4 and 5. This material is available free of charge via the 
Internet at http://pubs.acs.org.

\providecommand{\refin}[1]{\\ \textbf{Referenced in:} #1}

%\bibliographystyle{achemso}
%\bibliography{literature}

\providecommand{\refin}[1]{\\ \textbf{Referenced in:} #1}

%\AtEndDocument{
%\newpage
%\section*{For Table of Contents Use Only}
%
%New Benchmark Set of Coordination Reactions for the Assessment of Density Functionals
%
%{\bf Authors:} Thomas Weymuth, Erik P.~A.~Couzijn, Peter Chen, and Markus Reiher
%
%\begin{center}
%\includegraphics[scale=0.1]{toc.eps}
%\end{center}
%}

\end{document}